\documentclass[12pt]{article}
\usepackage{graphicx}
\usepackage{amsmath,amsfonts}
\newcommand{\sq}
{\nobreak \ifvmode \relax \else \ifdim\lastskip<1.5em
\hskip-\lastskip \hskip0.25em plus0em minus0.0em \fi \nobreak \vrule
height0.75em width0.75em depth0.0em\fi}
\usepackage{color}

\def\cC{\mbox{${\mathbb{C}}$}}

\begin{document}

\def\C{{\rm I\!\!\!\! C}}

\title{Achieving control of in-plane elastic waves}

\author{M. Brun$^*$, S. Guenneau$^{\dagger}$ \\and A.B.
Movchan$^{\dagger}$\\
\small{$^*$Department of Structural Engineering, University of
Cagliari,
Cagliari I-09123, Italy}\\
\small{$^\dagger$Department of Mathematical Sciences, Peach Street,
Liverpool, L69 3BX, UK }\\}

\maketitle

\begin{abstract}
We derive the elastic properties of a cylindrical cloak for in-plane
coupled shear and pressure waves. The cloak is characterized by a
rank 4 elasticity tensor with 16 spatially varying entries which are
deduced from a geometric transform.
Remarkably, the Navier equations retain their
form under this transform, which is generally untrue [Milton et al.,
New J. Phys. {\bf 8}, 248 (2006)]. We numerically check that clamped
and freely vibrating obstacles located inside the neutral region are
cloaked disrespectful of the frequency and the polarization of an
incoming elastic wave.

\end{abstract}
%

Recently, significant progress has been made on the control of
acoustic and electromagnetic waves. Transformation based solutions
to the conductivity and Maxwell's equations in curvilinear
coordinate systems, subsequently reported by Greenleaf {\it et al.}
\cite{greenleaf03} and then by Pendry {\it et al.} \cite{pendry06a}
and Leonhardt \cite{leonhardt06}, enable one to bend electromagnetic
waves around arbitrarily sized and shaped solids. More precisely,
the electromagnetic invisibility cloak is a metamaterial which maps
a concealment region into a surrounding shell: as a result of the
coordinate transformation the permittivity and permeability are
strongly heterogeneous and anisotropic within the cloak, yet
fulfilling impedance matching with the surrounding vacuum. The cloak
thus neither scatter waves nor induces a shadow in the transmitted
field. In \cite{pendry06b}, a cylindrical electromagnetic cloak
constructed using specially designed concentric arrays of split ring
resonators, was shown to conceal a copper cylinder around $8.5$ GHz.
The effectiveness of the transformation based cloak was numerically
demonstrated
solving Maxwell's equations using finite elements for an incident
plane wave (far field limit) \cite{cummer06a} and for electric line
current and magnetic loop sources (near field limit) \cite{zolla07}.
In \cite{cai07}, a reduced set of material parameters was introduced
to relax the constraint on the permeability, necessarily leading to
an impedance mismatch with vacuum which was shown to preserve the
cloak effectiveness to a good extent. Other routes to invisibility
include reduction of backscatter \cite{alu05} and cloaking through
anomalous localized resonances, the latter one using negative
refraction \cite{milton06a,milton07}. To date, a plethora of
research papers has been published in the fast growing field of
transformation optics.

However, transformation based invisibility cloaks applied to certain
types of elastodynamic waves in structural mechanics received less
attention, since the Navier equations do not usually retain their
form under geometric changes \cite{milton06b,norris}. One
simplification occurs for cylindrical geometries, whereby
out-of-plane shear waves decouple from in-plane waves. However,
in-plane shear and pressure waves remain inherently coupled. Earlier
proposals for neutral inclusions include using asymptotic and
computational methods to find suitable material parameters for
coated cylindrical inclusions \cite{bigoni98}. The latter has proved
successful in the elastostatic context in the case of anti-plane
shear and in-plane coupled pressure and shear polarizations.
However, neutrality breaks down for finite frequencies.

Other avenues to elastic cloaking should therefore be investigated.
For instance, Cummer and Schurig demonstrated that acoustic waves in
a fluid undergo the same geometric transform as electromagnetic
waves do and therefore retain their form \cite{cummer06b}. This
result has been since then generalized to three-dimensional acoustic
cloaks for pressure waves \cite{cummer08,chen07}. Importantly, such
cloaks require an anisotropic mass density which can be obtained via
a homogenization approach, which presents the advantage to be
broadband \cite{sanchez}. Acoustic cloaking for linear surface water
waves was chiefly achieved via the same mechanism in between $10$
and $15$ Hertz \cite{farhat08}.

In the present letter, we show that it is also possible to design a
cylindrical cloak for in-plane coupled pressure and shear elastic
waves. We demonstrate theoretically its unique mechanism and further
perform finite element computations checked again analytical
calculations of the Green's function for the Navier equations in
transformed coordinates. The main difference with previous work
\cite{milton06a} is that our elasticity tensor in the transformed
coordinates is no longer symmetric, which is a necessary condition
for the Navier equations to retain their form. Quite remarkably, we
find that the density remains a scalar quantity in the transformed
coordinates.

We consider the in-plane propagation of time-harmonic elastic waves
governed by the Navier equations
\begin{equation}
\nabla\cdot \cC:\nabla{\bf u} +\rho\,\omega^2 {\bf u}+{\bf b}={\bf
0} \; , \label{motion}
\end{equation}
where $\bf u$ is the displacement, $\rho$ the density, $\cC$ the
$4^{\mbox{th}}$-order constitutive tensor of the linear elastic
material and ${\bf b}={\bf b}({\bf x})$ represents the spatial
distribution of a simple harmonic body force $\hat{\bf b}({\bf
x},t)={\bf b}({\bf x})\exp(i\omega t)$, with $\omega$ the
wave-frequency and $t$ the time.

\begin{figure}
   \centerline{
\includegraphics[width=11cm]{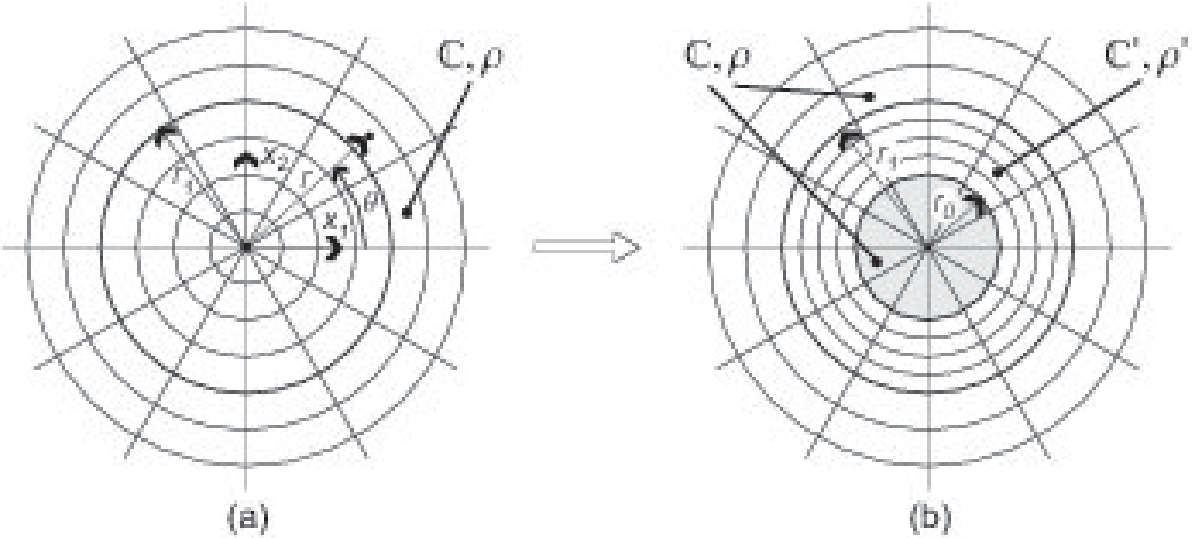}
   }
\vspace{-0.2cm} \caption{Geometric transform of
eqn.(\ref{PTransform}) from $(r,\theta)$ (Fig.\ref{figTransform}(a))
to $(r',\theta')$ (Fig.\ref{figTransform}(b)); $r_0$ and $r_1$ are
the inner and the outer radius of the cylindrical cloak,
respectively. The elastic constitutive tensor and the density in the
undeformed and in the deformed domains are denote by $\cC$, $\rho$
and $\cC'$, $\rho'$, respectively.} \label{figTransform}
\end{figure}

We introduce the geometric transform
$(r,\theta)\rightarrow(r',\theta')$ of \cite{greenleaf03,pendry06a}
\begin{equation}
\label{PTransform} \displaystyle \left\{
\begin{array}{lr}
r'=r_0+ \frac{r_1-r_0}{r_1} r \; , \; \theta'=\theta \;
,&\qquad\mbox{for}\,r\leq r1 \\
r'=r \; , \; \theta'=\theta \;
,&\qquad\mbox{for}\,r> r1 \\
\end{array}\right.
\end{equation}
shown in Fig. (\ref{figTransform}) and expressed in cylindrical
coordinates $r=\sqrt{x_1^2+x_2^2}$ and
$\theta=2\mbox{atan}(x_2/(x_1+\sqrt{x_1^2+x_2^2}))$, with $r_0$ and
$r_1$ the inner and outer radii of the circular cloak, respectively.

\begin{figure}[!h]
   \centerline{\includegraphics[width=13cm]{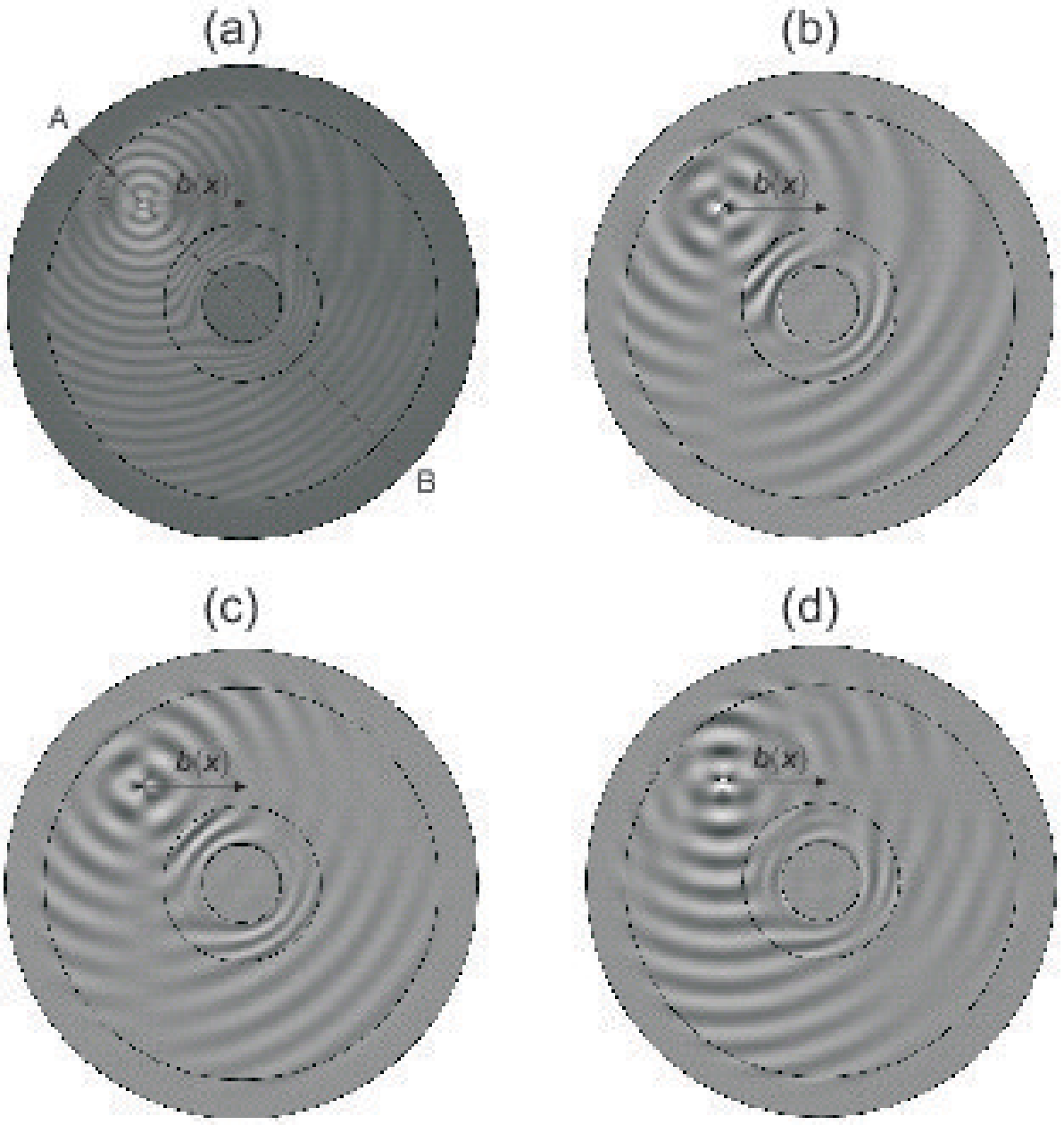}   }
\vspace{0.2cm} \caption{Elastic cloak in an elastic medium subjected
to a concentrated load. (a) Displacement magnitude
$u=\sqrt{u_1^2+u_2^2}$; (b) deformation
$\varepsilon_{11}=\frac{\partial u_1}{\partial x_1}$; (c)
deformation $\varepsilon_{22}=\frac{\partial u_2}{\partial x_2}$;
(d) deformation
$\varepsilon_{12}=\varepsilon_{21}=\frac{1}{2}(\frac{\partial
u_1}{\partial x_2}+\frac{\partial u_2}{\partial x_1})$.}
\label{fig1}
\end{figure}

By application of transformation (\ref{PTransform}), in the region
$r'\in [r_0,r_1]$ the Navier equations (\ref{motion}) are mapped
into the equations
\begin{equation}
\nabla\cdot{\cC'}:\nabla{\bf u}+\rho'\omega^2 {\bf u}={\bf 0} \;
,\label{snavier}
\end{equation}
where the body force is assumed to be zero, the stretched density
$\rho'=\frac{r-r_0}{r}(\frac{r_1}{r_1-r_0})^2\,\rho$ and elasticity
tensor $\cC'$ has non zero cylindrical components
\begin{equation}
\begin{array}{ll}
\cC'_{rrrr}=\frac{r-r_0}{r}(\lambda+2\mu), &
\cC'_{\theta\theta\theta\theta}=\frac{r}{r-r_0}(\lambda+2\mu),\\[1.6mm]
\cC'_{rr\theta\theta}=\cC'_{\theta\theta rr}=\lambda, &
\cC'_{r\theta\theta r}=\cC'_{\theta rr\theta}=\mu, \\[1.6mm]
\cC'_{r\theta r\theta}=\frac{r-r_0}{r} \mu, & \cC'_{\theta r\theta
r}=\frac{r}{r-r_0} \mu,
\end{array}
\label{sc}
\end{equation}
with $\lambda$ and $\mu$ the Lam\'e moduli characterizing the
isotropic behavior described by $\cC$.

Interestingly, the transformation (\ref{PTransform}) preserves the
isotropy of the density, which remains a scalar (yet spatially
varying) quantity in (\ref{snavier}), and avoids any coupling
between stress and velocity. This is a very unlikely situation for
elastodynamic waves propagating in anisotropic heterogeneous media
\cite{milton06b}. We also note that the proposed formulation poses
no limitations on the applied $\omega$ ranging form low to high
frequency, as the elasticity tensor does not depend upon $\omega$.


We report the finite element computations performed in the COMSOL
multiphysics package. The elastic cloak of equation (\ref{sc}) is
embedded in an isotropic elastic material with Lam\'e moduli
$\lambda=2.3$ and $\mu=1$ and density $\rho=1$, which are realistic
normalized parameters corresponding to fused silica. The elastic
cloak has inner and outer radii $r_0=0.2\, m$ and $r_1=0.4\, m$,
respectively.

\begin{figure}[!h]
   \centerline{\includegraphics[width=13cm]{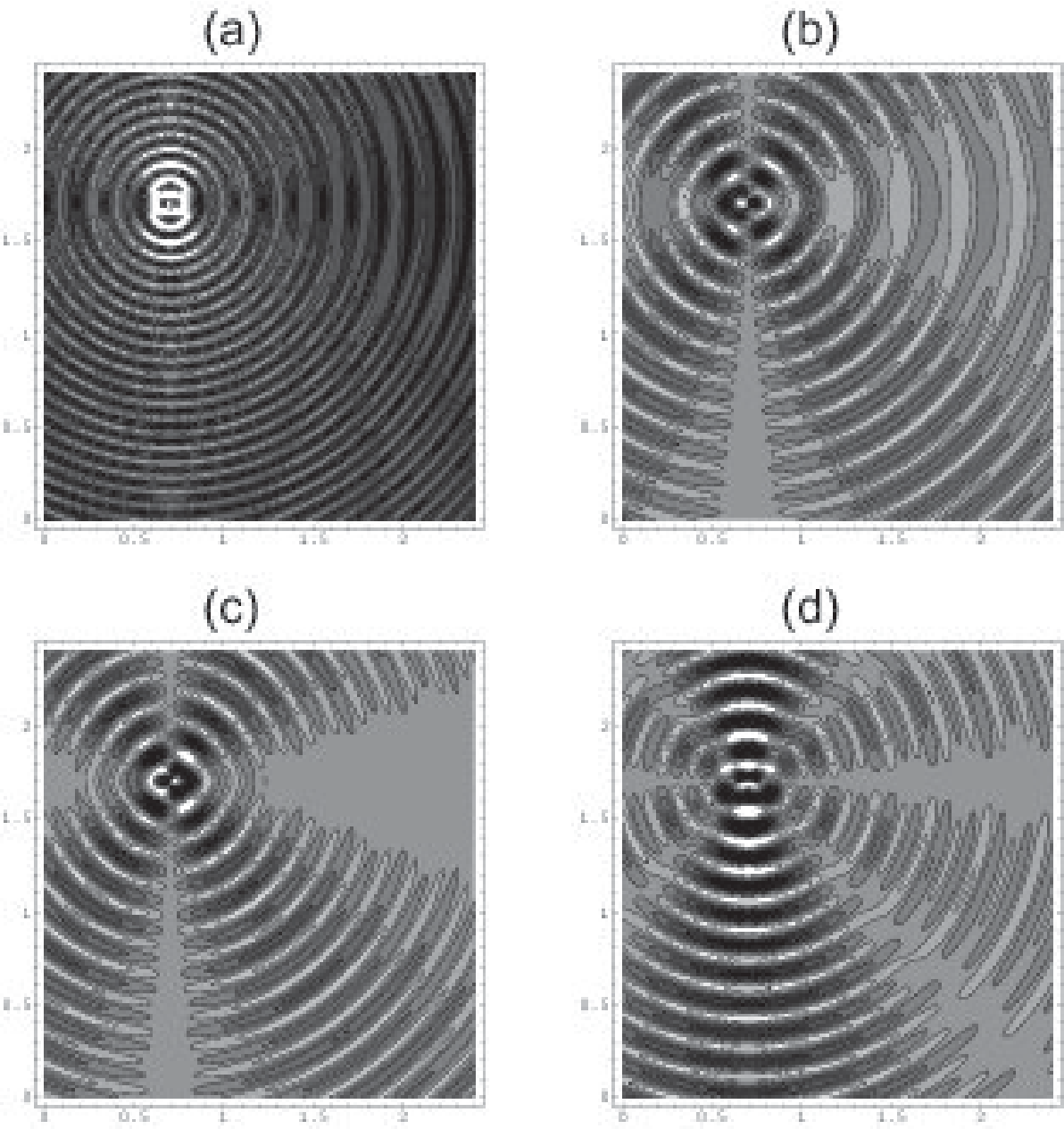}}
\vspace{0.cm} \caption{Harmonic Green's function in homogeneous
elastic space. (a) Displacement magnitude $u$; (b) deformation
$\varepsilon_{11}$; (c) deformation $\varepsilon_{22}$; (d)
deformation $\varepsilon_{12}$.} \label{fig2}
\end{figure}

The system is excited by an harmonic unit concentrated force applied
in direction $x_1$ and vibrating with angular frequency $\omega=40$
Hz.
A perfectly matched cylindrical layer has been
implemented in order to model the infinite elastic medium
surrounding the cloak (cf. outer ring on panels a, b, c and d of
Fig.\ref{fig1}); this has been obtained by application of the
geometric transform \cite{zh02},
\begin{equation}
r''=r_2+(1-i)(r-r_2), \qquad \theta''=\theta,
\end{equation}
where $r_2=1\,m$ is the inner radius of the outer ring in Fig.
(\ref{fig1}).

\begin{figure}[!h]
   \centerline{\includegraphics[width=13cm]{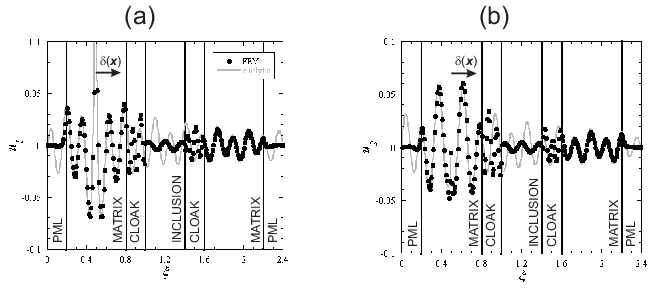}}
\vspace{0.cm} \caption{Comparison between numerical results in
presence of the elastic cloak of Fig. \ref{fig1} (black dots) and
Green's function in homogeneous elastic space of Fig. \ref{fig2}
(grey lines). Results are given along the line $AB$ detailed in Fig.
\ref{fig1}(a). (a) Horizontal displacement $u_1$; (b) Vertical
displacement $u_2$.} \label{fig3}
\end{figure}

In Fig. (\ref{fig1}) we clearly see that the wave patterns of the
displacement and deformations are smoothly bent around the central
region within the cloak (where the magnitudes are nearly zero).
Although the coupling of shear and pressure waves generated by the
concentrated force creates the optical illusion of interferences,
the comparison with the harmonic Green's function in homogeneous
elastic space (see \cite{WilMov95}) reported in Fig. (\ref{fig2})
shows, at least qualitatively, that there is neither forward nor
backward scattering. The absence of scattering is better detailed in
Fig. \ref{fig3} where results of Fig. (\ref{fig1}) and Fig.
(\ref{fig2}) are compared: the perfect agreement of the displacement
fields in the external matrix with and without the cloak is shown,
the distortion being bounded to the central region delimited by the
cloak. These are non-intuitive results, as the profiles of the
horizontal and vertical displacements in Fig. \ref{fig3} should
display a visible phase shift, since the associated acoustic paths
are different. More precisely, let us look at the expression of the
elasticity tensor given in (\ref{sc}). On the inner boundary of the
cloak, that is for $r=r_0$, its components $\cC'_{rrrr}$ and
$\cC'_{r\theta r\theta}$ vanish, whereas its components
$\cC'_{\theta\theta\theta\theta}$ and $\cC'_{\theta r\theta r}$ tend
to infinity. This physically means that pressure and shear waves
propagate with an infinite velocity in the $\theta$ direction along
the inner boundary, which results in a vanishing phase shift between
a wave propagating in a homogeneous elastic space and another one
propagating around the concealed region: this explains the
superimposed profiles of horizontal and vertical displacements in
Fig. \ref{fig3}.

In  conclusion, we have proposed an elastic cloak bending the
trajectory of in-plane coupled shear and pressure elastic waves
around a cylindrical obstacle. The cloak can be designed by the use
of heterogeneous density and heterogeneous and anisotropic elastic
stiffness; the distribution of the physical properties has been
obtained with the introduction of stretched coordinates. Our results
open new vistas in cloaking devices for elastodynamic waves in
anisotropic media, yet with an isotropic density.

Part of this work was performed while MB was working under the EPSRC
grant EP/F027125/1.



\end{document}